\documentclass[preprint,preprintnumbers,amsmath,amssymb]{revtex4}

\usepackage{bm}
\usepackage{graphicx}
\usepackage{subfig}
\usepackage{array}
\usepackage{color}

\graphicspath{{figures/}}

\makeatletter

\newcommand{\Rmnum}[1]{\expandafter\@slowromancap\romannumeral #1@}
\makeatother

\begin{document}

\title{On the breaking of $\mu$-$\tau$ flavor symmetry}

\author{Zhen-hua Zhao } \email{zhaozhenhua@ihep.ac.cn}

\affiliation{Department of Physics, Liaoning Normal University, Dalian 116029, China \\
Institute of High Energy Physics, Chinese Academy of Sciences, Beijing 100049, China}

\begin{abstract}
In light of the observation of a relatively large $\theta^{}_{13}$, one has to consider
breaking the $\mu$-$\tau$ symmetry properly which would otherwise result in a
vanishing $\theta^{}_{13}$ (as well as $\theta^{}_{23} = \pi/4$).
Therefore, we investigate various symmetry-breaking patterns
and accordingly identify those that are phenomenologically viable.
Furthermore, the symmetry-breaking effects arising from some specific physics
(e.g., the renormalization group equation running effect) are discussed as well \cite{review}.
\end{abstract}

\maketitle

\section{Introduction}

As acknowledged by the 2015 Nobel prize in physics, the discovery of neutrino oscillations \cite{PDG}
implies that neutrinos are massive as opposed to the standard model (SM) setting. So far the most
plausible way of generating small neutrino masses has been the seesaw mechanism \cite{seesaw} which
leads neutrinos to be the Majorana particles. We therefore take on this possibility and deal with
the symmetric neutrino mass matrix $M^{}_\nu$. Consequently, the neutrino mixing may arise from
the mismatch between their mass and flavor eigenstates \cite{PMNS}.
Such a mixing is described by one $3 \times 3$ unitary matrix $U = U^\dagger_l U^{}_\nu$
with $U^{}_\nu$ and $U^{}_l$ being respectively the unitary matrix for diagonalizing $M^{}_\nu$ and
the charged-lepton mass matrix $M^{}_l M^\dagger_l$. In the standard parametrization, $U$ is expressed as
\begin{equation}
U=P^{}_\phi
\left( \begin{matrix}
c_{12}c_{13} & s_{12}c_{13} & s_{13}e^{-{\rm i} \delta} \cr
-s_{12}c_{23}-c_{12}s_{23}s_{13}e^{{\rm i}\delta} & c_{12}c_{23}-s_{12}s_{23}s_{13}e^{{\rm i}\delta}  & s_{23}c_{13} \cr
s_{12}s_{23}-c_{12}c_{23}s_{13}e^{{\rm i}\delta}  & -c_{12}s_{23}-s_{12}c_{23}s_{13}e^{{\rm i}\delta} & c_{23}c_{13}
\end{matrix} \right) P^{}_\nu \;,
\label{1}
\end{equation}
with the definition $c^{}_{ij} = \cos{\theta^{}_{ij} }$ and $s^{}_{ij} = \sin {\theta^{}_{ij}}$
(for $ij=12,13,23$). The diagonal phase matrix $P^{}_\nu = {\rm Diag}(e^{{\rm i} \rho}, e^{{\rm i} \sigma}, 1)$
contains two Majorana CP phases, whereas $P^{}_{\phi} = {\rm Diag}(e^{{\rm i} \phi^{}_1}, e^{{\rm i} \phi^{}_2},
e^{{\rm i} \phi^{}_3})$ consists of three unphysical phases that can be removed by the charged-lepton
field redefinitions.

The mixing angles $\theta^{}_{ij}$ as well as the Dirac CP phase $\delta$ can be measured in the
neutrino-oscillation experiments \cite{PDG}. It is interesting to find that $\theta^{}_{23}$
has a value close to $\pi/4$ from the atmospheric neutrino-oscillation experiment \cite{SuperK}.
In comparison, $\theta^{}_{13}$ has not been determined until very recently \cite{DYB}
and was then only constrained by the upper limit $\sin^2_{}{2\theta^{}_{13}}<0.18$ \cite{Chooz}.
Given these experimental facts, one was naturally tempted to take $\theta^{}_{23}=\pi/4$ together with
$\theta^{}_{13}=0$ as an ideal possibility which in turn motivated intensive studies about the
$\mu$-$\tau$ symmetry \cite{MT}. This symmetry is defined in a way that $M^{}_l$ is diagonal while
$M^{}_\nu$ keeps invariant under the transformation $\nu^{}_\mu \leftrightarrow \nu^{}_\tau$ and
thus takes a form as
\begin{equation}
M^{}_\nu= \left( \begin{matrix}
A & \hspace{0.1cm} B & \hspace{0.1cm} B \cr
B & \hspace{0.1cm} C & \hspace{0.1cm} D \cr
B & \hspace{0.1cm} D & \hspace{0.1cm} C
\end{matrix} \right) \;.
\end{equation}
It is straightforward to show that we will have $\theta^{}_{23}=\pi/4$ and $\theta^{}_{13}=0$
as a result of these symmetry conditions. Note that the $\mu$-$\tau$ symmetry has no definite prediction
for $\theta^{}_{12}$. Nevertheless, a further condition $A + B = C + D$ imposed on the $M^{}_\nu$
given by Eq. (2) will give $\sin{\theta^{}_{12}}=1/\sqrt{3}$ in which case we are left with
the ever-popular tri-bimaximal (TB) mixing pattern \cite{TB}.

However, the experimental results (i.e., the observed $\theta^{}_{13} \simeq 0.15$ \cite{DYB} as well as
a possible deviation of $\theta^{}_{23}$ from $\pi/4$)
go against this simple flavor symmetry \cite{data}:
\begin{eqnarray}
&& \sin^2{\theta^{}_{13}} = 0.0215 - 0.0259 \;, \hspace{0.5cm}
 \Delta m^2_{21}= (7.32 -7.80) \times 10^{-5} {\rm eV}^2 \;, \nonumber \\
&& \sin^2{\theta^{}_{23}} = 0.414 - 0.594  \;, \hspace{0.65cm}
|\Delta m^2_{32}| \simeq |\Delta m^2_{31}| = (2.32 -2.49) \times 10^{-3} {\rm eV}^2 \;.
\end{eqnarray}
Here the results for neutrino mass squared differences $\Delta m^2_{ij} = m^2_i - m^2_j$ (for $ij=21,32,31$)
are also presented for later use. Note that the sign of $\Delta m^2_{32}$ (equivalently $\Delta m^2_{31}$)
has not been determined yet, leaving us with two possibilities for the neutrino mass ordering
(i.e., $m^{}_1 < m^{}_2 < m^{}_3$ or $m^{}_3 < m^{}_1 < m^{}_2$).
In addition, the absolute neutrino mass scale remains unknown as well.
Although the $\mu$-$\tau$ symmetry must be broken to accommodate these experimental results \cite{he},
it may still be taken as a starting point for understanding the neutrino mixing pattern if in some situations
this symmetry holds to a good approximation.
In the next section we will study what kind of approximately $\mu$-$\tau$ symmetric $M^{}_\nu$
can lead to phenomenologically viable results.
While section 3 is devoted to a further discussion about the symmetry-breaking effects induced by
some specific physics (e.g., the renormalization group equation (RGE) running effects).
Finally, we summarize our main results in the last section.

\section{A general study for the symmetry-breaking effects}

Since the mixing matrix is derived from the mass matrix, we prefer to discuss the symmetry-breaking
effects at the mass matrix level. In order to measure the symmetry-breaking strength, we introduce
two dimensionless quantities \cite{grimus}
\begin{eqnarray}
\epsilon^{}_{1}=\frac{M^{}_{e \mu}- M^{}_{e\tau}}{M^{}_{e \mu}+
M^{}_{e\tau}} \ ,\hspace{0.7cm} \epsilon^{}_{2}=\frac{M^{}_{\mu
\mu}-M^{}_{\tau \tau}}{M^{}_{\mu \mu}+ M^{}_{\tau\tau}} \;,
\end{eqnarray}
which correspond to the defining features of $\mu$-$\tau$ symmetry
(i.e., $M^{}_{e\mu} = M^{}_{e\tau}$ and $M^{}_{\mu\mu} = M^{}_{\tau\tau}$).
By virtue of these two quantities, the most general neutrino mass matrix
can always be parameterized into the form
\begin{eqnarray}
M^{}_{\nu}=\left( \begin{matrix}
 A & \hspace{0.3cm}  B \left(1+\epsilon^{}_1\right) & \hspace{0.3cm} B \left(1-\epsilon^{}_1\right) \cr
B \left(1+\epsilon^{}_1\right) & \hspace{0.3cm} C \left(1+\epsilon^{}_2\right) & \hspace{0.3cm}  D \cr
B \left(1-\epsilon^{}_1\right) & \hspace{0.3cm} D  & \hspace{0.3cm} C \left(1-\epsilon^{}_2\right)
\end{matrix} \right) \;.
\end{eqnarray}
When $|\epsilon^{}_{1,2}|$ are simultaneously small enough (e.g., $<0.2$),
one can argue that $M^{}_\nu$ assumes an approximate $\mu$-$\tau$ symmetry.
Instead of acquiring the possible values of $|\epsilon^{}_{1,2}|$ via the reconstruction of
$M^{}_\nu$ in terms of the neutrino masses and mixing matrix $U$ \cite{GJP},
we start from an approximately $\mu$-$\tau$ symmetric $M^{}_\nu$ (i.e.,
$|\epsilon^{}_{1,2}|$ are assumed to be small in the first place)
and explore its implications for $\theta^{}_{13}$ and $\Delta \theta^{}_{23}$.
According to the naturalness argument, the sizes of $\theta^{}_{13}$ and $\Delta \theta^{}_{23}$
will be directly controlled by $\epsilon^{}_{1,2}$.
By making perturbation expansions for the small parameters in diagonalizing
the $M^{}_\nu$ given by Eq. (5), one will arrive at the following relations connecting
$\theta^{}_{13}$ and $\Delta \theta^{}_{23} \equiv \theta^{}_{23}- \pi/4$ to $\epsilon^{}_{1,2}$
\cite{grimus}:
\begin{eqnarray}
& \theta^{}_{13} e^{-{\rm i} \delta}  & = ( 2\Delta m^{2}_{31} )^{-1}
[ 2m^{}_3 m^{}_{12} c^2_{12} \epsilon^{}_1 + 2\overline m^{}_1
m^*_{12} c^2_{12} \epsilon^*_1 + m^{}_3 (m^{}_{22} + m^{}_3)
c^{}_{12} s^{}_{12} \epsilon^{}_2 \nonumber \\
& & + \overline m^{}_1 (m^*_{22} + m^{}_3 ) c^{}_{12} s^{}_{12} \epsilon^*_2 ]
+ ( 2\Delta m^{2}_{32} )^{-1} [ 2 m^{}_3 m^{}_{12} s^2_{12}\epsilon^{}_1
+2 \overline m^{}_2 m^*_{12} s^2_{12} \epsilon^*_1 \nonumber \\
& & -m^{}_3 ( m^{}_{22} + m^{}_3 ) c^{}_{12} s^{}_{12} \epsilon^{}_2 -
\overline m^{}_2 (m^*_{22} + m^{}_3 )c^{}_{12} s^{}_{12}\epsilon^*_2 ] \;,
\nonumber \\
 & \hspace{-0.4cm}  \Delta\theta^{}_{23} & \hspace{-0.4cm} =  {\rm Re} \{ (2\Delta m^{2}_{31} )^{-1}
[ 2m^{}_{12} c^{}_{12} s^{}_{12} (\overline m^{*}_{1} \epsilon^{}_1 + m^{}_3
\epsilon^{*}_{1} ) + (m^{}_{22}+ m^{}_3 ) s^2_{12} (\overline
m^{*}_{1} \epsilon^{}_2 + m^{}_3 \epsilon^{*}_{2} ) ]
\nonumber \\
 & \hspace{-0.4cm} & \hspace{-0.4cm} - ( 2\Delta m^{2}_{32} )^{-1}
[ 2m^{}_{12} c^{}_{12} s^{}_{12} (\overline m^{*}_{2} \epsilon^{}_1 +
m^{}_3 \epsilon^{*}_{1} )  - (m^{}_{22} + m^{}_3 ) c^2_{12}
(\overline m^{*}_{2} \epsilon^{}_2 + m^{}_3 \epsilon^{*}_{2} ) ] \}  \;, \hspace{0.45cm}
\end{eqnarray}
where we have defined
\begin{eqnarray}
m^{}_{11} =  \overline m^{}_{1}c^{2}_{12} +
\overline m^{}_{2}s^{2}_{12} \;, \hspace{0.3cm}
m^{}_{12} = \left(\overline m^{}_{1} - \overline m^{}_{2}\right)
c^{}_{12} s^{}_{12} \;, \hspace{0.3cm}
m^{}_{22}  =  \overline m^{}_{1} s^{2}_{12} + \overline
m^{}_{2}c^{2}_{12} \;,
\end{eqnarray}
with $\overline{m}^{}_1 \equiv m^{}_1 e^{2 {\rm i} \rho}$ and
$\overline{m}^{}_2 \equiv m^{}_2 e^{2 {\rm i} \sigma}$.
With the help of these results, one can study
the dependence of $\theta^{}_{13} e^{-{\rm i} \delta}$ and $\Delta\theta^{}_{23}$
on $\epsilon^{}_{1,2}$ in some special situations to be given below.

First of all, let us work under the assumption of CP conservation in which case Eq. (6) is
reduced to
\begin{eqnarray}
\theta^{}_{13} = \frac{2 m^{}_{12} c^2_{12}\epsilon^{}_1
+ (m^{}_{22} + m^{}_3) c^{}_{12}s^{}_{12}\epsilon^{}_2}
{2 \left(m^{}_3 \mp m^{}_1\right)}
 + \frac{2m^{}_{12} s^2_{12}
\epsilon^{}_1 - (m^{}_{22} + m^{}_3) c^{}_{12} s^{}_{12}\epsilon^{}_2}
{2  \left(m^{}_3 \mp m^{}_2\right)} \; ,
\nonumber \\
\Delta\theta^{}_{23} = \frac{2m^{}_{12} c^{}_{12}
s^{}_{12}\epsilon^{}_1 + (m^{}_{22} + m^{}_3) s^2_{12}\epsilon^{}_2}
{2 \left(m^{}_3 \mp m^{}_1\right)}
- \frac{2m^{}_{12}
c^{}_{12}s^{}_{12}\epsilon^{}_1 - (m^{}_{22} + m^{}_3) c^2_{12}\epsilon^{}_2}
{2  \left(m^{}_3 \mp m^{}_2\right)} \;,
\end{eqnarray}
where $\mp$ correspond to $\overline {m}^{}_1 = \pm m^{}_1$ (and $\overline {m}^{}_2 = \pm m^{}_2$).
It is found that the values of $\theta^{}_{13}$ and $\Delta \theta^{}_{23}$ are strongly
dependent on the neutrino mass spectrum as well as the Majorana phases once the symmetry-breaking
strength is specified. (1) When $m^{}_1$ is vanishingly small, $\theta^{}_{13}$ is well approximated by
\begin{eqnarray}
\theta^{}_{13} \sim \frac{1}{2}\sqrt{\frac{\Delta m^2_{21}}{\Delta m^2_{31}}} \
c^{}_{12}s^{}_{12}\left(2\epsilon^{}_{1}-\epsilon^{}_{2}\right)
\simeq 0.04 \left(2\epsilon^{}_{1}-\epsilon^{}_{2}\right ) \;,
\end{eqnarray}
which, given $|\epsilon^{}_{1,2}|<0.2$, is definitely unacceptable.
(2) For $m^{}_{1}\simeq m^{}_2 \gg m^{}_3$, the results will depend on
the combination of $\rho$ and $\sigma$. When
$\rho$ is equal to $\sigma$, $\theta^{}_{13}$ is extremely suppressed as shown by
\begin{eqnarray}
\theta^{}_{13}\sim \frac{1}{4} \frac{\Delta m^2_{21}}{ \Delta m^2_{31}} c^{}_{12}s^{}_{12}
\left(2\epsilon^{}_1 -\epsilon^{}_2\right) \simeq -0.004 \left(2\epsilon^{}_1-\epsilon^{}_2
\right) \;.
\end{eqnarray}
Otherwise, $\theta^{}_{13}$ approximates to
\begin{eqnarray}
\theta^{}_{13}\sim \frac{1}{2} \cos{2\theta^{}_{12}}
\sin{2\theta^{}_{12}} \left(2{\epsilon}^{}_1-\epsilon^{}_{2}\right)
\simeq 0.18 \left(2{\epsilon}^{}_1-\epsilon^{}_{2}\right) \;,
\end{eqnarray}
which is still unable to give the observed value.
(3) When neutrinos assume a nearly degenerate mass spectrum $m^{}_1 \simeq m^{}_2 \simeq m^{}_3$
and $(\rho, \sigma) = (0, 0)$, one obtains $\theta^{}_{13}$ as
\begin{eqnarray}
\theta^{}_{13} \sim \frac{2m^{2}_1}{\Delta m^2_{31}} \frac{\Delta m^2_{21}}{ \Delta m^2_{31}}
c^{}_{12} s^{}_{12}\epsilon^{}_2 \;,
\end{eqnarray}
which is at most 0.03 by taking account the constraint $m^{}_1 + m^{}_2 + m^{}_3 < 0.23$ eV from
cosmological observations \cite{planck}.
In the case of $(\rho, \sigma) = (0, \pi/2)$, one will have
\begin{eqnarray}
\theta^{}_{13}  \sim   \frac{2m^2_1}{\Delta m^2_{31}} c^{}_{12}s^{}_{12}
\left(2c^2_{12}\epsilon^{}_{1}+s^2_{12}\epsilon^{}_{2} \right) \;, \hspace{0.5cm}
\Delta\theta^{}_{23}  \sim  \frac{2m^2_1}{\Delta m^2_{31}}s^{2}_{12}
\left(2c^2_{12}\epsilon^{}_{1}+s^2_{12}\epsilon^{}_{2}\right) \;.
\end{eqnarray}
Thanks to the enhancement factor $m^2_1/\Delta m^2_{31}$, $\theta^{}_{13}$ can easily reach
the observed value. Noteworthy, the correlation between $\theta^{}_{13}$ and $\Delta \theta^{}_{23}$
will give the prediction $|\Delta\theta^{}_{23}|\sim \theta^{}_{13} s^{}_{12}/c^{}_{12} \simeq6^\circ$
which can be tested by precision measurements for $\theta^{}_{23}$.
Finally, it turns out that the cases of $(\rho, \sigma) = (\pi/2, 0)$ and $(\pi/2, \pi/2)$ are not
capable of generating a realistic $\theta^{}_{13}$ or $\Delta \theta^{}_{23}$.

When CP violation is concerned, more interesting possibilities will arise.
In the first example we assume $\rho$ and $\sigma$ to be 0 or $\pi/2$ and $\epsilon^{}_{1,2}$ to be purely
imaginary (parameterized as $\epsilon^{}_{1,2} = {\rm i}|\epsilon^{}_{1,2}|$).
One immediately from Eq. (7) obtains $\Delta \theta^{}_{23} =0$,
$\delta =\pm \pi/2$ and
\begin{eqnarray}
\theta^{}_{13} =  \frac{2m^{}_{12} c^2_{12}
|\epsilon^{}_1| + (m^{}_{22} + m^{}_3) c^{}_{12}s^{}_{12}|\epsilon^{}_2|} {2
\left(m^{}_3 \pm m^{}_1\right)}
+ \frac{2m^{}_{12} s^2_{12}|\epsilon^{}_1|
-(m^{}_{22} + m^{}_3) c^{}_{12}s^{}_{12}|\epsilon^{}_2|}
{2 \left(m^{}_3 \pm m^{}_2\right)} \;.
\end{eqnarray}
A similar analysis as in the CP conservation case shows that
the observed $\theta^{}_{13}$ is only obtainable under the condition of a nearly degenerate
neutrino mass spectrum in combination with $(\rho, \sigma)= (0, \pi/2)$ or $(\pi/2, 0)$.
At this point it is worth mentioning that these results (i.e., trivial Majorana phases,
maximal Dirac phase and $\theta^{}_{23}=\pi/4$)
are the same as those predicted by an $M^{}_\nu$ respecting the $\mu$-$\tau$
reflection symmetry \cite{reflection}. Such an interesting symmetry is defined as follows:
in the basis where $M^{}_l$ is diagonal, $M^{}_\nu$ should keep invariant with respect to
the transformation
\begin{eqnarray}
\nu^{}_e \to \nu^c_e \;, \hspace{0.5cm} \nu^{}_\mu \to \nu^c_\tau \;,
\hspace{0.5cm} \nu^{}_\tau \to \nu^c_\mu \;,
\end{eqnarray}
and thus appears as
\begin{eqnarray}
M^{}_{\nu} = \left( \begin{matrix}
A & B & B^{*} \\ B & C & D \\ B^* & D & C^*
\end{matrix} \right) \;,
\end{eqnarray}
with $A$ and $D$ being real parameters. It is easy to check that the $M^{}_\nu$ given by Eq. (5)
happens to acquire this symmetry in the symmetry-breaking scenario under discussion.
In another example, we instead assume $\rho$ and $\sigma$ to take non-trivial values
and $\epsilon^{}_{1,2}$ to be real. From Eq. (7) one can see that a finite $\delta$ may arise
from the non-trivial Majorana phases even when $\epsilon^{}_{1,2}$ themselves are real \cite{mei}.
And its magnitude is not directly controlled by the symmetry-breaking parameters.
This is easy to understand from that the $\mu$-$\tau$ symmetry has no power of
constraining the value of $\delta$. For illustration, $\delta$ will be given by
\begin{eqnarray}
\tan{\delta}= \frac{m^{}_2 \sin{2\sigma}-m^{}_1\sin{2\rho}} {m^{}_1
\cos{2\rho}-m^{}_2 \cos{2\sigma}-m^{}_3 \Delta m^2_{21}/\Delta m^2_{31} } \;,
\end{eqnarray}
in the special case of $\epsilon^{}_2 = 2 \epsilon^{}_1$ which as one can see later resembles
the symmetry breaking induced by the RGE running effect.

So far the symmetry-breaking terms have been supposed to be relatively small
as compared with the entry itself they reside in. But we will relax this constraint
when dealing with an $M^{}_\nu$ with a hierarchical structure.
For an $M^{}_\nu$ of this kind, one can assume that the dominant entries emerge at
the leading order (LO) while the sub-dominant entries become finite only after receiving the
next-to-leading-order (NLO) contributions which may also perturb the dominant entries.
If the LO and NLO contributions are assumed to keep and break the $\mu$-$\tau$ symmetry
respectively, the sub-dominant entries will be completely occupied
by the symmetry-breaking terms. This speculation motivates us to reconsider the physical
implications of an approximately $\mu$-$\tau$ symmetric $M^{}_\nu$. A good example in this
regard is one hierarchical neutrino mass matrix that will lead to $m^{}_1 < m^{}_2
\ll m^{}_3$. It may be parameterized in a form as \cite{mohapatra}
\begin{eqnarray}
M^{}_{\nu}= m \left( \begin{matrix} d \epsilon & c\epsilon & b\epsilon \\
c\epsilon & 1+a\epsilon & -1 \\ b\epsilon & -1 & 1+\epsilon
\end{matrix} \right) \;,
\end{eqnarray}
where $\epsilon$ is a small quantity used to characterize the relative size
of the NLO contributions compared to the LO ones, while $a,b,c$ and $d$ are
$\mathcal O(1)$ real coefficients. This neutrino mass matrix leads us to the
mixing angles
\begin{eqnarray}
\theta^{}_{13} \simeq \frac{1}{2\sqrt 2}\left(b-c\right) \epsilon \;,
\hspace{0.5cm}
\Delta \theta^{}_{23} \simeq \frac{1}{4}\left(a-1\right) \epsilon \;,
\end{eqnarray}
and mass eigenvalues
\begin{eqnarray}
m^{}_{1,2} \simeq \frac{1}{4} \epsilon m \left(2d+a+1 \mp \Delta \right) \;,
\hspace{0.5cm} m^{}_{3} \simeq 2m \;,
\end{eqnarray}
with $\Delta=\sqrt{\left(2d-a-1\right)^2+8\left(b+c\right)^2}$.
By fitting these mass eigenvalues
with the measured $\Delta m^2_{21}$ and $\Delta m^2_{32}$, one finds $\epsilon \sim
\sqrt{\Delta m^2_{21}/\Delta m^2_{32}} \simeq 0.15$. Hence the smallness of $\theta^{}_{13}$
finds an explanation from the hierarchy between $\Delta m^2_{21}$ and $\Delta m^2_{32}$
in this particular scenario. Another example is the neutrino mass matrix
\begin{eqnarray}
M^{}_{\nu}= m \left( \begin{matrix}
e\epsilon & 1+d\epsilon & 1+c\epsilon \\
1+d\epsilon & b\epsilon & \epsilon \\
1+c\epsilon & \epsilon & a\epsilon
\end{matrix} \right) \;,
\end{eqnarray}
that results in $m^{}_1 = - m^{}_2$ and $m^{}_3 = 0$ at the LO. Note that the LO terms respect the well-known
$L^{}_e - L^{}_\mu - L^{}_\tau$ symmetry \cite{STP} (with $L$ standing for the lepton number) while the NLO terms violate it.
A straightforward calculation yields the mixing angles
\begin{eqnarray}
\theta^{}_{13} \simeq \frac{\left(a-b\right) \epsilon}{2\sqrt 2} \; ,
\hspace{0.5cm}
\Delta \theta^{}_{23}  \simeq \frac{\left(c-d\right) \epsilon}{2} \; ,
\end{eqnarray}
and mass eigenvalues
\begin{eqnarray}
m^{}_{1,2} \simeq \frac{1}{4}\left[\left(2e+2+a+b \right)\epsilon \mp
4\sqrt{2}\right] m \; ,
\hspace{0.5cm}
m^{}_3 \simeq \frac{1}{2}\left(a+b-2\right) \epsilon m \;.
\end{eqnarray}
Fitting these mass eigenvalues with $\Delta m^2_{21}$ and $\Delta m^2_{32}$ requires $\epsilon$ to be at
the order of $\Delta m^2_{21}/ \Delta m^2_{31}$, implying that $\theta^{}_{13}$ would be exceedingly suppressed.
Hence this pattern of $M^{}_\nu$ is disfavored by the current experimental data.

\section{Symmetry breaking arising from some specific physics}

In this section we study the symmetry-breaking effects arising from some specific physics.
Above all, it should be noted that the RGE running effect may break the $\mu$-$\tau$ symmetry.
From the phenomenological point of view, flavor symmetries are usually implemented at a superhigh
energy scale $\Lambda^{}_{\rm FS}$ so as to keep away from the low-energy constraints.
One should therefore take into account this effect when confronting the physical
consequences of a flavor symmetry with the experimental data available at low energies
$\Lambda^{}_{\rm EW}$ \cite{rge}. In the RGE running process the significant difference between $m^{}_\mu$
and $m^{}_\tau$ will perform as a natural source for the symmetry breaking. In the minimal
supersymmetry standard model (MSSM), the running of $M^{}_\nu$ is governed by \cite{rges}
\begin{eqnarray}
\frac{{\rm d}M^{}_{\nu}}{{\rm d}t} =  \left(Y^{\dagger}_l Y^{}_l \right)^{T} M^{}_\nu +
M^{}_\nu \left(Y^{\dagger}_l Y^{}_l \right) + \alpha M^{}_{\nu} \;,
\end{eqnarray}
where $\alpha \simeq -6/5 g^2_1-6 g^2_2 +6 y^2_t$ and
$Y^{}_l = {\rm Diag} (y^{}_e, y^{}_\mu, y^{}_\tau)$ denote the Yukawa couplings
for charged leptons among which $y^{}_e$ and $y^{}_\mu$ will be neglected in the following discussions.
The reason for us to work in the MSSM is that the value of $y^2_\tau = (1 + \tan^2{\beta} )
m^2_\tau/v^2$ may be greatly enhanced by choosing a large $\tan{\beta}$.
A $\mu$-$\tau$ symmetric $M^{}_\nu$ at $\Lambda^{}_{\rm FS}$ can be expressed in terms of the
corresponding physical quantities in a form as
\begin{eqnarray}
M^{}_{\nu}=\left( \begin{matrix} \vspace{0.15cm}
m^{}_{11} & \hspace{0.2cm} -\displaystyle\frac{1}{\sqrt{2}} m^{}_{12} &  \hspace{0.2cm} -\displaystyle
\frac{1}{\sqrt{2} }m^{}_{12} \\ \vspace{0.15cm}
 \cdots & \hspace{0.2cm} \displaystyle \frac{1}{2}(m^{}_{22} +m^{}_{3})
& \hspace{0.2cm} \displaystyle \frac{1}{2}(m^{}_{22}-m^{}_{3}) \\
\cdots & \hspace{0.2cm} \cdots & \hspace{0.2cm} \displaystyle \frac{1}{2}( m^{}_{22}+m^{}_{3})
\end{matrix} \right) \;.
\end{eqnarray}
By integrating Eq. (24) one obtains the neutrino mass matrix at $\Lambda^{}_{\rm EW}$ as \cite{irge}
\begin{eqnarray}
M^{\prime}_{\nu} & = & I^{}_{\alpha} {\rm Diag}(1,1,1-\Delta^{}_{\tau})
M^{}_{\nu} {\rm Diag}(1,1,1-\Delta^{}_{\tau})  \nonumber \\
& = & I^{}_{\alpha}
\left( \begin{matrix} \vspace{0.15cm} m^{}_{11}
&- \displaystyle \frac{1}{\sqrt{2}}m^{\prime}_{12} (1+ \displaystyle \frac{1}{2}\Delta^{}_\tau )&
-\displaystyle \frac{1}{\sqrt{2}}m^{\prime}_{12}(1- \displaystyle \frac{1}{2}\Delta^{}_\tau) \\
\vspace{0.15cm} \cdots & \displaystyle \frac{1}{2}(m^{\prime}_{22}
+m^{\prime}_{3})(1+\Delta^{}_\tau)
 & \displaystyle \frac{1}{2}(m^{\prime}_{22}-m^{\prime}_{3}) \\
\cdots & \cdots & \displaystyle \frac{1}{2}(
m^{\prime}_{22}+m^{\prime}_{3})(1-\Delta^{}_\tau)
\end{matrix} \right) \;,
\end{eqnarray}
with
\begin{eqnarray}
I^{}_{\alpha} = {\rm exp} \left(\frac{1}{16\pi^2}
\int^{\Lambda^{}_{\rm EW}}_{\Lambda^{}_{\rm FS}} \alpha {\rm d}t
\right) \;, \hspace{0.5cm}
\Delta^{}_{\tau}=\frac{1}{16\pi^2}\int^{\Lambda^{}_{\rm
FS}}_{\Lambda^{}_{\rm EW}}y^2_{\tau} {\rm d}t \;,
\end{eqnarray}
and
\begin{eqnarray}
m^{\prime}_{12}=m^{}_{12}(1-\frac{1}{2}\Delta^{}_{\tau})\; ,
\hspace{0.5cm} m^{\prime}_{22}=m^{}_{22}(1-\Delta^{}_{\tau})\; ,
\hspace{0.5cm} m^{\prime}_{3}=m^{}_{3}(1-\Delta^{}_{\tau})\; .
\end{eqnarray}
Numerically, for $\Lambda^{}_{\rm FS} =10^{14}$ GeV, $I^{}_\alpha$ and $\Delta^{}_\tau$
respectively range from 0.9 to 0.8 and from 0.002 to 0.044 when $\tan{\beta}$ varies
from 10 to 50.
The physical quantities at $\Lambda^{}_{\rm EW}$ can be extracted by diagonalizing $M^{\prime}_\nu$
with a unitary matrix $U^\prime$. After a straightforward calculation
one finds the mixing angles \cite{DGR}
\begin{eqnarray}
&& \theta^{\prime}_{12} \simeq  \theta^{}_{12}+
\frac{1}{2}c^{}_{12}s^{}_{12} \frac{\left|\overline m^{}_1+\overline
m^{}_2\right|^2}{\Delta m^2_{21}} \Delta^{}_{\tau} \; ,
\hspace{0.5cm}
\theta^{\prime}_{13} \simeq  c^{}_{12}s^{}_{12}
\frac{m^{}_{3}\left| \overline m^{}_1 - \overline
m^{}_2 \right|} {\Delta m^2_{31}} \Delta^{}_{\tau} \; ,
\nonumber \\
&& \theta^{\prime}_{23} \simeq  \frac{\pi}{4}+
\frac{\left|\overline m^{}_1 + m^{}_3\right|^2s^{2}_{12} +
\left|\overline m^{}_2+m^{}_3\right|^2c^{2}_{12}}{2 \Delta m^2_{31}}
\Delta^{}_{\tau} \;.
\end{eqnarray}
From these results we can draw the following conclusions concerning the running behaviours of
$\theta^{}_{13}$ and $\theta^{}_{23}$: Even when the absolute neutrino mass
scale reaches its upper limit from cosmological observations, $\tan{\beta}$ still should be
larger than 50 in order to generate a realistic $\theta^{}_{13}$ \cite{GPRR}. However, such a
$\tan{\beta}$ would be problematic by rendering the bottom-quark Yukawa coupling
non-perturbatively large \cite{MSSM}. It is thus fair to say that the observed
$\theta^{}_{13}$ can not be purely generated from the radiative effects \cite{jue}.
As for $\theta^{}_{23}$, an appreciable deviation of it from $\pi/4$
can be acquired when the neutrino mass spectrum is nearly degenerate. Interestingly,
this deviation will be positive (negative) in the case of $\Delta m^2_{31}
>0$ ($<0$), providing a potential correlation between the octant of $\theta^{}_{23}$
and the neutrino mass ordering \cite{LZ}.

In the above discussions $M^{}_l$ has been taken to be diagonal. When this is not the case,
the unitary matrix $U^{}_l$ will also contribute to the neutrino mixing
according to $U= U^\dagger_l U^{}_\nu$ \cite{correction}. Such a contribution may become relevant
when a certain texture of $M^{}_\nu$ fails to give viable phenomenological
consequences or $M^{}_l$ is constrained to be non-diagonal by some physics (e.g.,
the connection between $M^{}_l$ with the mass matrix for down-type quarks in
the grand unified theory (GUT) models). If $U^{}_\nu$ results from
a $\mu$-$\tau$ symmetric $M^{}_\nu$, $U^{}_l$ may bring about the deviations of $\theta^{}_{13}$ and
$\theta^{}_{23}$ from 0 and $\pi/4$. So let us explore the physical implications of $\mu$-$\tau$ symmetry
breaking from the charged lepton sector. To make things easier,
a slightly different parametrization for the $3 \times 3$ unitary matrix from the standard one
will be adopted:
\begin{eqnarray}
U  =  U^{}_{23} U^{}_{13} U^{}_{12} P^{}_{\alpha}
 = \left( \begin{matrix} 1 & 0 & 0 \\ 0 & c^{}_{23} &
\tilde{s}^{*}_{23} \\ 0 & -\tilde{s}^{}_{23} & c^{}_{23} \end{matrix} \right)
\left( \begin{matrix} c^{}_{12} & \tilde{s}^{*}_{12} & 0 \\
-\tilde{s}^{}_{12} & c^{}_{12} & 0 \\ 0 & 0 & 1  \end{matrix} \right)
\left( \begin{matrix} c^{}_{13} & 0 & \tilde{s}^{*}_{13} \\ 0 & 1 &
0 \\ -\tilde{s}^{}_{13} & 0 & c^{}_{13} \end{matrix} \right) P^{}_\alpha \;,
\end{eqnarray}
with $\tilde s^{}_{ij} = s^{}_{ij} e^{ {\rm i} \delta^{}_{ij}}$ (for $ij = 12, 13, 23$) and
$P^{}_{\alpha} = {\rm Diag} (e^{ {\rm i} \alpha^{}_{1}}, e^{ {\rm i} \alpha^{}_{2}}, e^{ {\rm i} \alpha^{}_{3}})$.
This new parametrization is related to the standard one via the phase transformations
\begin{eqnarray}
& & \delta^{}_{12}=\phi^{}_2-\phi^{}_1 \; , \hspace{0.5cm}
\delta^{}_{13}=\delta+\phi^{}_3-\phi^{}_1 \; ,
\hspace{0.5cm} \delta^{}_{23}=\phi^{}_3-\phi^{}_2\; ,
\nonumber \\
& & \alpha^{}_1=\phi^{}_1+\rho \; ,\hspace{0.75cm}
\alpha^{}_2=\phi^{}_2+\sigma \; , \hspace{1.35cm} \alpha^{}_3=\phi^{}_3 \;.
\end{eqnarray}
Correspondingly, the neutrino mixing will be obtained as
\begin{eqnarray}
U  =  U^{\dagger}_{l}U^{}_{\nu} = P^{l\dagger}_{\alpha}
U^{l\dagger}_{12} U^{l\dagger}_{13} U^{l\dagger}_{23}
U^{\nu}_{23}U^{\nu}_{13}U^{\nu}_{12}P^{\nu}_{\alpha} =
P^{l\dagger}_{\alpha}U^{}_{23}U^{}_{13}U^{}_{12}P^{\nu}_{\alpha} \;.
\end{eqnarray}
Here we concentrate on the case of $U^{}_l$
being approximately diagonal for two considerations: the $U^{}_\nu$ resulting from a $\mu$-$\tau$
symmetric $M^{}_\nu$ is already close to the realistic $U$, so the corrections
from $U^{}_l$ need not be too significant; by analogy with the quark sector, an approximately
diagonal $U^{}_l$ is expected as a natural outcome in light of the large mass hierarchies among
charged leptons.
As a result, the mixing angles in $U$ approximate to \cite{AK}
\begin{eqnarray}
\tilde s^{}_{13} \simeq  -\tilde \theta^{l}_{13} c^{\nu}_{23}
-\tilde \theta^{l}_{12}\tilde s^{\nu}_{23} \;, \hspace{0.5cm}
\tilde s^{}_{12} \simeq \tilde s^{\nu}_{12}-\tilde \theta^{l}_{12} c^{\nu}_{12}c^{\nu}_{23} +\tilde
\theta^{l}_{13}c^{\nu}_{12}\tilde s^{\nu*}_{23} \;,
\end{eqnarray}
with $\theta^{\nu}_{23}=\pi/4$.
These results can be further simplified by assuming $\theta^{l}_{13} \ll \theta^{l}_{12}$:
\begin{eqnarray}
\delta \simeq \delta^l_{12} - \delta^\nu_{12} - \pi \;,
\hspace{0.5cm}
\theta^{}_{13} \simeq \theta^{l}_{12} s^{\nu}_{23} \; ,
\hspace{0.5cm}
s^{}_{12} \simeq s^{\nu}_{12} +
\theta^{}_{13}c^{\nu}_{12} \cos{\delta} \;.
\end{eqnarray}
If $\theta^{l}_{12}$ has a value close to the Cabibbo angle of quark
mixing $\theta^{}_{\rm C} \simeq 0.22$, then the second expression in Eq. (34) becomes
$\theta^{}_{13} \simeq \theta^{}_{\rm C}/\sqrt{2}$ which agrees well with
the observations. This remarkable relation makes the idea of relating the lepton and quark sectors
in the GUT models particularly attractive \cite{gut}. Moreover, the last expression in Eq. (34)
implies a correlation between $\theta^{\nu}_{12}$ and $\delta$. For instance, if $\theta^{\nu}_{12}$
has a value as in the TB mixing pattern (i.e., $\sin{\theta^{\nu}_{12}}= 1/\sqrt{3}$ which is close
to the real $\theta^{}_{12}$), $\delta$ should lie around $\pm \pi/2$ so that the contribution from
the second term to $\theta^{}_{12}$ can be suppressed.

Finally, we point out that the mixing between active and sterile neutrinos can serve as another source for
the symmetry breaking. Sterile neutrinos, as the name suggests, do not carry any quantum
number under the SM gauge symmetry and thus do not take part in the SM interactions. Although there has not been direct
evidence for sterile neutrinos, their existence is either theoretically motivated or experimentally
hinted. A good example on the theoretical side is the heavy right-handed neutrino introduced for
implementing the seesaw mechanism. On the experimental side, the long-standing LSND anomaly \cite{LSND}
and several other short-baseline neutrino-oscillation anomalies \cite{short} imply the possible existence of
an $\mathcal O(\rm eV)$ sterile neutrino which mixes with the active neutrinos. It is therefore worthwhile
for us to investigate the implementation of $\mu$-$\tau$ symmetry in the presence of sterile neutrinos.
One interesting possibility in this connection is just that sterile neutrinos may be responsible for the
symmetry breaking \cite{sterile}. To be specific, in the 3+1 neutrino mixing scheme (i.e.,
three active neutrinos plus one sterile neutrino), the $4 \times 4$ neutrino mass matrix can be
parameterized as
\begin{eqnarray}
M = \left( \begin{matrix} m^{}_{ee} & m^{}_{e\mu} & m^{}_{e\mu} & m^{}_{e
s} \cr m^{}_{e\mu} & m^{}_{\mu\mu} & m^{}_{\mu\tau} & m^{}_{\mu s}
\cr m^{}_{e\mu} & m^{}_{\mu\tau} & m^{}_{\mu\mu} & m^{}_{\tau s} \cr
m^{}_{e s} & m^{}_{\mu s} & m^{}_{\tau s} & m^{}_{s s}
\end{matrix} \right)  \;.
\end{eqnarray}
Note that the upper-left $3\times 3$ sub-matrix has been assumed to keep the $\mu$-$\tau$ symmetry,
while $m^{}_{\mu s} \neq m^{}_{\tau s}$ will be taken as the source for symmetry breaking.
This would be a reasonable assumption when sterile neutrinos have a different mass origin from
the active neutrinos so that the formers do not necessarily respect the symmetry possessed by
the latters.

\section{Summary}

In summary, we have performed a systematic study on the various $\mu$-$\tau$ symmetry breaking patterns
in order to accommodate the observed $\theta^{}_{13}$ in a proper way.
In the first approach, two parameters $\epsilon^{}_{1,2}$ are introduced to characterize the
symmetry breaking and required to be small (e.g., $|\epsilon^{}_{1,2}| <0.2$)
so as to keep the symmetry as an approximate one.
When CP is conserved, an approximately $\mu$-$\tau$ symmetric $M^{}_\nu$ is capable of producing a viable
$\theta^{}_{13}$ only under the condition of a nearly degenerate neutrino mass spectrum in combination
with $(\rho, \sigma) = (0, \pi/2)$ in which case a $|\Delta \theta^{}_{23}| \simeq 6^\circ$ is also predicted.
In the particular case that $\epsilon^{}_{1,2}$ are purely imaginary while $\rho$ and $\sigma$ take
trivial values, one is led to $\Delta \theta^{}_{23} =0$ and $\delta = \pm \pi/2$ as predicted by
the $\mu$-$\tau$ reflection symmetry. When an $M^{}_\nu$ with a hierarchical structure is concerned,
another approach may be invoked: the LO effects which respect the symmetry only contribute to the
dominant entries, while the symmetry-breaking NLO effects are responsible for generating the sub-dominant
entries as well as perturbing the dominant ones. One hierarchical $M^{}_\nu$ that leads to
$m^{}_1 < m^{}_2 \ll m^{}_3$ turns out to be a good illustration for this approach of implementing
the approximate $\mu$-$\tau$ symmetry.

On the other hand, some specific physics that may give rise to the $\mu$-$\tau$ symmetry breaking
have been discussed as well. First of all, the RGE running effect always
serves as a source for the symmetry breaking when this symmetry is implemented at
an energy scale much higher than $\Lambda^{}_{\rm EW}$. However, this effect is not sufficient for generating
the observed $\theta^{}_{13}$ from 0 even in the optimal situation where the absolute neutrino mass
scale and $\tan{\beta}$ take their largest allowed values. Furthermore, when $M^{}_l$ is
not diagonal for some reasons, $U^{}_l$ will also contribute to the neutrino mixing. If $U^{}_l$ features
$\theta^{l}_{12} \simeq \theta^{}_{\rm C} \gg \theta^{l}_{13}$ and $U^{}_\nu$ results from an $M^{}_\nu$
respecting the $\mu$-$\tau$ symmetry, then an interesting relation
$\theta^{}_{13} \simeq \theta^{}_{\rm C}/ \sqrt{2}$ is reached which has strengthened the motivation for
relating the quark and lepton sectors. Last but not least, the mixing between active and sterile
neutrinos (whose existence is hinted by a few short-baseline neutrino-oscillation anomalies)
may also be responsible for the breaking of this interesting symmetry.

\section*{Acknowledgements}

I would like to thank Professor Z. Z. Xing for fruitful collaboration on the
$\mu$-$\tau$ flavor symmetry. I am also grateful to H. Fritzsch and K. K. Phua for their warm
hospitality at the IAS-NTU where the Conference on New Physics at the Large Hadron Collider was held.
This work was supported in part by the China Postdoctoral Science Foundation under grant
No. 2015M570150 and by the National Natural Science Foundation of China under Nos. 11135009
and 11375207.

\bibliographystyle{ws-rv-van}

\end{document}